\documentclass[twocolumn,aps,prl,amssymb,superscriptaddress,showpacs]{revtex4-1}
\usepackage[latin9]{inputenc}
\usepackage{verbatim}
\usepackage{amsmath}
\usepackage{amssymb}
\usepackage{graphicx}

\makeatletter
\usepackage{amsfonts}
\usepackage{dsfont}
\usepackage{bm}
\usepackage{appendix}
\usepackage{appendix}

\def\tr{\mathrm{tr}}

\usepackage[usenames,dvipsnames]{color}
\usepackage{ulem}
\usepackage{soul}
\setstcolor{red}
\sethlcolor{yellow}
\makeatother

\begin{document}

\title{Ubiquitous non-local entanglement with Majorana zero modes}

\author{Alessandro Romito}

\affiliation{%
\mbox{Department of Physics, Lancaster University, Lancaster LA1 4YB, United Kingdom}}

\author{Yuval Gefen}
\affiliation{%
\mbox{Department of Condensed Matter Physics, The Weizmann Institute of Science, Rehovot 76100, Israel}}
\begin{abstract}
Entanglement in quantum mechanics contradicts local realism, and is a manifestation of quantum non-locality. Its presence can be detected through the violation of Bell, or CHSH inequalities. Paradigmatic quantum systems provide examples of both, non-entangled and entangled states. Here we consider a minimal complexity setup consisting of 6 Majorana zero modes. We find that {\it any} allowed state in the degenerate Majorana space is non-locally entangled. We show how to measure (with available techniques) the CHSH-violating correlations, using either intermediate strength or weak measurement protocols.
\end{abstract}
\maketitle

\paragraph{Introduction}

Majorana zero-modes are particular non-Abelian quasi-particles that reflect the topologically non-trivial character of the underlying system. Over less than a decade Majorana zero modes (MZM) have crossed the line from  mathematically intriguing solid state manifestations of Majorana's original particles \cite{Moore1991,Read1999,Kitaev2001}, to experimentally realizable entities \cite{Lutchyn2010,Oreg2010,Wilczek2009}. Being a class of non-Abelian anyons \cite{Wilczek1982,Stern2010}, MZM offer a paradigm for fault-tolerant information processing \cite{Nayak2008}. Following initial experiments \cite{Mourik2012,Das2012,Churchill2013,Nadj-Perge2014,Albrecht2016,Deng2016}, we are now at the stage where specific platforms for engineering and manipulating Majoranas \cite{aasen2015,Plugge2016,Plugge2016a} are being implemented.  It is broadly felt that implementations of topological states of matter for quantum information processing should rely, first, on thorough understanding of quantum states defined by Majorana zero modes. Interestingly, a unique property of MZM is that they may constitute a manifestation of quantum non-locality. Indirect observable signatures emerging from non-local MZM (albeit not a proof of their non-locality) have been studied earlier in setups based on mesoscopic superconductors \cite{Fu2010,Michaeli2016,Vijay2016}, or coupled Majorana zero modes \cite{Rosenow2012,Rubbert2016}. 

Non-locality is an indispensable pillar of quantum mechanics. For a system made of at least two particles non-locality is a manifestation of quantum entanglement between spatially distinct degrees of freedom. For paradigmatic systems, an apt example being two spin-1/2 particles, it is possible to construct both entangled (e.g. a singlet) and non-entangled, i.e. product (e.g. triplet-1) states \cite{footnote1}. Quantum non-locality is quantified by Bell's inequality  \cite{Bell1965}, or, in a manner that is more conducive to experimental testing \cite{Giustina2015,Hensen2015}, by  the violation of the Clauser, Horne, Shimony and Holt (CHSH) inequality \cite{Clauser1969}. Entanglement properties of Majorana systems have been explored as a source of non-locality \cite{Campbell2014}, for applications to certifiable random numbers generation \cite{Deng2013}, and for extending protected operations beyond braiding \cite{Clarke2015}.

The focus of the present study is the direct observability of distinct entanglement features of quantum states in the degenerate space defined by Majorana zero modes. We identify a system of minimal complexity (minimal number of MZM). For that system: (i) we show that {\it any} allowed quantum state in the degenerate space defined by  a set of MZM is non-locally entangled; (ii) we then demonstrate how such entanglement can be detected within technologically feasible measurement platforms; (iii) finally, we show how our entanglement detection protocol can be realized within weak measurement operations.

\paragraph{Model}

\begin{figure}
\protect\includegraphics[width=0.85\columnwidth]{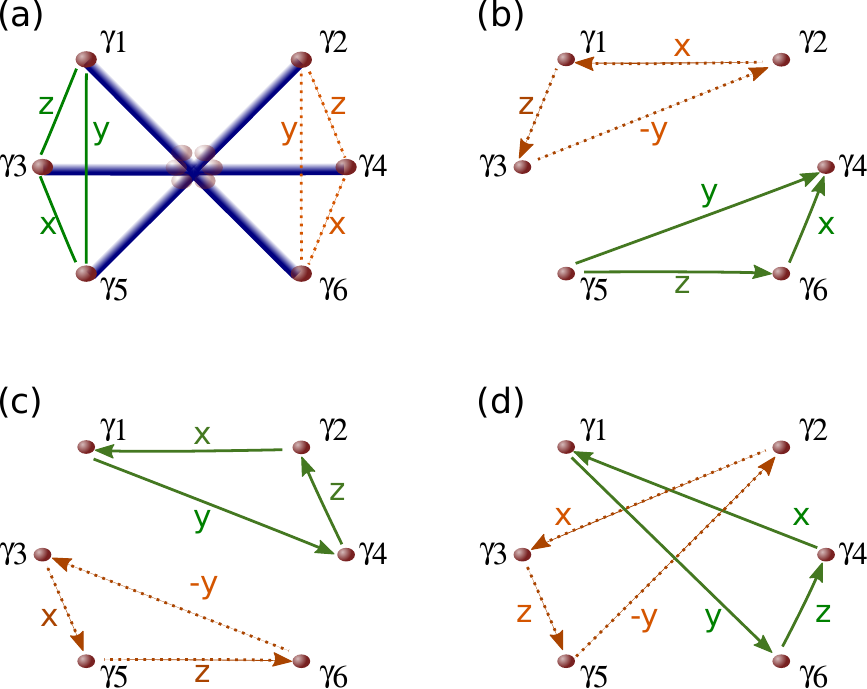}
\caption{(a): Multi-terminal junction of topological superconducting wires (blue) hosting Majorana zero modes at their ends (red dots).
The Majorna end-states at the junction (fading red) are gapped up and the
six Majorana end-states (solid red dots) $\gamma_1, \dots, \gamma_6$ constitute the low energy excitations of the system. A left detector (not shown) measures the operators indicated by
solid (green) arrows, which are labeled by the corresponding spin algebra operators,
e.g. $Z=-i\gamma_{1}\gamma_{2} $ playing the role of $\sigma_z$. Dashed (orange) arrows define the operators measured by the right detector. Panels (a-d) show different possible partitionings into left and right sectors with the corresponding operators. CHSH inequalities are necessarily violated in at least one of the partitionings in (a-d).
\label{Fig:system} }
\end{figure}

Our system consists of a multi-terminal junction made up of an even number of one-dimensional topological superconductors (branches), depicted in  Fig. \ref{Fig:system}(a). They all have a common end point at the center. 
Such junctions can be engineered experimentally with semiconductor wires \cite{Mourik2012,Das2012,Churchill2013,Albrecht2016} or magnetic impurity chains \cite{Nadj-Perge2014}.
Each segment, $\alpha \in \{ 1,\dots, 2N\}$, of this setup consists of  a 1-d spinless p-wave superconductor characterized by a bulk excitation energy gap, $\Delta_{\alpha}$, and by zero-energy MZM at the end points, $\gamma_{\alpha}$, $\gamma' _{\alpha}$,  localized at the wire's boundaries. The dynamics of these MZM is underlined by the algebra $\{\gamma_{\alpha},\gamma{}_{\beta}\}=2\delta_{\alpha,\beta}$,
$\{\gamma_{\alpha},\gamma'{}_{\beta}\}=0$. The wire Hamiltonian
at energies well below the  gap is given by $H_{\alpha}=\epsilon_{\alpha}\gamma_{\alpha}\gamma'_{\alpha}$,
where $\epsilon_{\alpha}\sim e^{-l_{\alpha}/\xi_{\alpha}}$ is exponentially small with the wire's length, $l_\alpha$. The latter is larger than the superconductor coherence length \cite{Kitaev2001,Lutchyn2010,Oreg2010} $\xi_{\alpha}\propto1/\Delta_{\alpha}$. We also assume $\epsilon_\alpha=0$, which is a valid assumption as long as the duration of the measurement protocol is not too long.
 At the junction,
the Josephson coupling between each pair of branches, $\alpha,$$\beta$
results in a low energy coupling between the corresponding Majorana
end-states. This is described by the tunneling Hamiltonian \cite{Kitaev2001,Alicea2011} $H_{T}=\sum_{\alpha,\beta}t_{\alpha,\beta}\gamma'_{\alpha}\gamma'_{\beta}$, which, generically, pairs up the $2N$  Majorana  zero modes, $\gamma'_1, \dots, \gamma'_{2N}$,
to finite energy states  with energies $\sim \min \{t_{\alpha,\beta}\}$. These states are then projected out of the degenerate ground-state space. The MZM $\{\gamma_{\alpha} \}$ far from the junction (see Fig. \ref{Fig:system}(a)) represent the remaining zero energy degrees of freedom, which span a $2^{N}$ degenerate ground state. The Majorana subspace does not accommodate a well-defined number of fermions: it may exchange pairs of fermions with the underlying superconductor. It follows that the parity of the Majorana system, $\mathcal{P}=i\prod_{j=1}^{{2N}}\gamma_{j}$,
is a good quantum number, hence the degenerate ground-state space consists of
two subspaces, each  of a definite parity. Without loss of generality, we may restrict ourselves, in the low-energy space, to states in the $2^{N-1}$-dimensional odd subspace. We will study a minimal complexity setup consisting of $2N=6$ MZM.

The MZM, $\gamma_1, \dots, \gamma_{2N}$, can be partitioned into two different sets, left ($L$), and right
($R$);  each of which is to be probed by a separate external detector.
The detectors can be tuned to measure any combination of pairs of
Majorana products. Physically this is a measurement of the occupancy
of certain Dirac fermions degrees of freedom, constructed  from the Majorana degrees
of freedom. Details of the measurement procedure are discussed below. 
An example to be utilized below is depicted
in Fig. \ref{Fig:system}(a), where the $L$-set consists of
$\gamma_{1}$, $\gamma_{3}$, $\gamma_{5}$, and the coupled detector
can measure any operator of the form
\begin{equation}
\hat{O}_{{\rm L}}=-i(\cos\theta_{L}\gamma_{1}\gamma_{3}+\sin\theta_{L}\cos\phi_{L}\gamma_{3}\gamma_{5}+\sin\theta_{L}\sin\phi_{L}\gamma_{5}\gamma_{1}).\label{eq:operator}
\end{equation}
 Note that the expectation values of the measured observables are bounded,
${-1} \leqslant \langle O_{L}\rangle\leqslant 1$ (the eigenvalues of the bilinear Majorana products are $\pm 1$). Genuine quantum correlations underlying a state can be identified through the expectation values of correlated
measurements. Specifically, a state that can be described within a local hidden variable theory  (a.k.a. local realism), satisfies 
the CHSH inequality \cite{Clauser1969}
\begin{equation}
\mathcal{C\equiv}\vert\langle \hat{O}_{L} \hat{O}_{R}\rangle-\langle \hat{O}_{L} \hat{O}'_{R}\rangle\vert+\vert\langle \hat{O}'_{L} \hat{O}'_{R}\rangle+\langle \hat{O}'_{L} \hat{O}_{R}\rangle\vert\leqslant2,\label{eq:CHSH}
\end{equation}
where $\hat{O}_L$, $\hat{O}'_{L}$ and $\hat{O}_{R}$, $\hat{O}'_{R}$ 
are pairs of spatially
separable sets of observables. For quantum non-locally entangled states, it is
instead possible to choose the operators such that  \cite{Cirelson1980} 
$2<\mathcal{C}\leqslant2\sqrt{2}$, hence providing evidence of genuine quantum correlations.
Eq.\eqref{eq:CHSH} is an equivalent formulation of Bell's inequality \cite{Clauser1969,Bell1965}, which, relying only on averaged correlation outputs, can
be tested directly by averaging over repeated measurements, including weak measurements.

A quantum system generically realizes both entangled states that violate Bell's (hence CHSH) inequalities and product states. The novel aspect of our work is that we show that in the degenerate space spanned by  Majorana zero modes, 
{\it any} state is non-locally entangled. In other words, one can always find (at least) one partitioning of the  MZM into two
spatially separable sets, where the CHSH inequality is violated. Loosely speaking,
for any state in the degenerate ground-space it is possible to design non-local measurements that reveal intrinsic non-locality.  

To begin with, we realize that establishing the CHSH relations requires measurement of 
non-commuting observables for each of the separated-in-space parts of the system, {\it i.e.} the $L$ and $R$ sets. For observables bilinear in the elementary MZM (cf.  Eq. \eqref{eq:operator}),
this requires a minimum of three Majorana operators for each set. The minimal complexity 
setup appropriate for our purpose is therefore 
 a multi-terminal junction consisting of six branches (6 MZM)
 (cf. Fig. \ref{Fig:system}(a)). We  consider a
 generic state of the 4-degenerate odd parity ground-manifold.
Following the labeling of the Majoranas  in Fig.\ref{Fig:system}(a), such a state
is parametrized as
\begin{equation}
\vert\psi\rangle=A\,d_{1,3}^{\dagger}d_{4,2}^{\dagger}d_{5,6}^{\dagger}\vert0\rangle+B\,d_{1,3}^{\dagger}\vert0\rangle+C\,d_{4,2}^{\dagger}\vert0\rangle-D\,d_{5,6}^{\dagger}\vert0\rangle,\label{eq:state}
\end{equation}
where $\vert A\vert^2 + \vert B\vert^2 +\vert C\vert^2 +\vert D\vert^2 =1$. Here we have introduced the fermionic degrees of freedom, $d_{1,3}^{\dagger}=(\gamma_{1}+i\gamma_{3})/2$
$d_{4,2}^{\dagger}=(\gamma_{4}+i\gamma_{2})/2$, $d_{5,6}^{\dagger}=(\gamma_{5}+i\gamma_{6})/2$,
and the state $\vert0\rangle$ is defined by $d_{1,3}\vert0\rangle=0$,
$d_{5,6}\vert0\rangle=0$, $d_{4,2}\vert0\rangle=0$. Throughout our analysis we will switch between  Fock space states, spin-{1/2} states, and Majorana notation.  

Consider the partitioning depicted in Fig. \ref{Fig:system}(a):  $\gamma_{1}$, $\gamma_{3}$, $\gamma_{5}$
constitute the  ($L$) set; $\gamma_{2}$, $\gamma_{4}$, $\gamma_{6}$ --
the  ($R$) set.  The operators $\hat{Z}_{L}\equiv-i\gamma_{1}\gamma_{3}$,
$\hat{X}_{L}\equiv-i\gamma_{3}\gamma_{5}$, $\hat{Y}_{L}\equiv-i\gamma_{5}\gamma_{1}$
satisfy the  Pauli matrice algebra,  $\sigma_{z}=\hat{Z}_L$, $\sigma_{x}=\hat{X}_L$,
$\sigma_{y}=\hat{Y}_L$. It follows that measurement of an operator of the form
of Eq. (\ref{eq:operator}) can be mapped onto the measurement
of $\hat{O}_L=\hat{\bm{\sigma}} \cdot \mathbf{n}$, where $\hat{\bm{\sigma}}=2 \hat{\bm{S}}$ is a  spin-$1/2$ operator and $\mathbf{n}\equiv(\sin \theta_L \cos \phi_L,\sin \theta_L \sin \phi_L, \cos \theta_L)$ .
Analogously, $\hat{Z}_{R}\equiv-i\gamma_{4}\gamma_{2}$, $\hat{Y}_{R}\equiv i\gamma_{2}\gamma_{6}$,
$\hat{X}_{R}\equiv-i\gamma_{6}\gamma_{4}$ can be identified with Pauli
operators of the right set. In such spin-$1/2$ language the state
$\vert\psi\rangle$ reads $\vert\psi\rangle=A\vert\uparrow_{L}\uparrow_{R}\rangle+B\vert\uparrow_{L}\downarrow_{R}\rangle+C\vert\downarrow_{L}\uparrow_{R}\rangle+D\vert\downarrow_{L}\downarrow_{R}\rangle$, where $\vert\uparrow_{i} \rangle$ $\vert \downarrow_{i}\rangle$ are the eigenstates of $\hat{Z}_i$ ($i=L,R$).
The maximal value of the CHSH correlation $\mathcal{C}$ in Eq. \ref{eq:CHSH}
is given by
\begin{equation}
\mathcal{C}_{135|246}=2\sqrt{1+4\left|AD-BC\right|^{2}},\label{eq:quantify-1}
\end{equation}
where the subscript indicates the partitioning in which the measurment is performed.
For any state, $ 2 \leqslant \mathcal{C}_{135|246} \leqslant 2 \sqrt{2}$, and $\mathcal{C}_{135|246} \neq 2$ signals non-local correlations, which happens unless
$AD-BC=0$. Operationally, this means that, if $AD-BC\neq0$ one can select the  coefficients
$\theta_{L}$, $\theta_{R}$, $\phi_{L},$ $\phi_{R}$ to construct a proper set
 of operators that violate the CHSH inequality.

Though a given state might not violate the CHSH inequality with measurements within the specific $L$ and $R$ sets,  
it can still lead to a violation of the CHSH inequality with a different partitioning of the MZM. 
The new partitioning will be non-local in the old $L$ and $R$ sets.
For example, a different partitioning consisting of the sets $\tilde{L}$,
and $\tilde{R}$ is depicted in Fig.  \ref{Fig:system}(b), where the left detector is connected
to $\gamma_{5}$, $\gamma_{6}$, $\gamma_{4}$ while $\gamma_{1}$,
$\gamma_{3}$, $\gamma_{2}$ are connected to the right detector.
In this case we define he operators $\hat{Z}_{\tilde{L}}\equiv-i\gamma_{5}\gamma_{6}$,
$\hat{X}_{\tilde{L}}\equiv-i\gamma_{6}\gamma_{4}$, $\hat{Y}_{\tilde{L}}\equiv-i\gamma_{4}\gamma_{5}$,
and $\hat{Z}_{\tilde{R}}\equiv-i\gamma_{1}\gamma_{3}$, $\hat{Y}_{\tilde{L}}\equiv i\gamma_{3}\gamma_{2}$,
$\hat{X}_{\tilde{L}}\equiv-i\gamma_{2}\gamma_{1}$. Mapping the problem to that of 
a two spin $1/2$ system, we can write the state $\vert\psi\rangle$
in Eq. (\ref{eq:state}) as $\vert\psi\rangle=\tilde{A}\vert \tilde{\uparrow}_{L} \tilde{\uparrow}_{R}\rangle + \tilde{B} \vert \tilde{\uparrow}_{L}\tilde{\downarrow}_{R}\rangle + \tilde{C} \vert \tilde{\downarrow}_{L} \tilde{\uparrow}_{R}\rangle + \tilde{D} \vert \tilde{\downarrow}_{L}\tilde{\downarrow}_{R}\rangle$, where $\tilde{A}=A$, $\tilde{B}=D$, $\tilde{C}=B$, $\tilde{D}=C$, and $\vert \tilde{\uparrow}_i \rangle$, $\vert \tilde{\downarrow}_i\rangle$ are the eigenstates of $\tilde{Z}_i$.
For the given $A$, $B$, $C$, $D$, the maximal violation of the CHSH inequalities in the new partitioning (maximal with respect of the choice of $\hat{O}_L$, $\hat{O}_R$, cf. Eq. \eqref{eq:operator}) is given by 
\begin{equation}
\mathcal{C}_{564|132}=2\sqrt{1+4\left|AC-DB\right|^{2}},\label{eq:quantify-1-1}
\end{equation}
where $\mathcal{C}_{135|246}>2$ signals non-local correlations. This is achieved 
unless $AC-BD=0$. Measurements of CHSH inequalities following the partitionings depicted in panels (c) and (d) of Fig. \ref{Fig:system} yield $\mathcal{C}_{421|563}=2\sqrt{1+4\left|AB-CD\right|^{2}}$
and $\mathcal{C}_{641|352}=2\sqrt{1+\left|A^2+C^2+D^2-B^2\right|^{2}}$. The
condition $\mathcal{C}_{135|246}=\mathcal{C}_{136|245}=\mathcal{C}_{456|123}=\mathcal{C}_{124|356} = 2$ 
can never be fulfilled, i.e. CHSH correlations will be non-local in
at least one of the four partitions considered in Fig. \ref{Fig:system}.

It is important at this point to make the following observation. The 
partitioning of the MZM into two sets naturally leads to the definition of operators
satisfying spin-$1/2$ algebra for each set. Different partitionings entails different sets of 
spin operators. 
Such a construction of operators is not unique for MZM. It can be done 
for any quantum system whose state is spanned in a 4-dimensional space. 
Consider the case of two real, physical,  spin-$1/2$ degrees of freedom,  associated with $L$ and $R$ respectively, which are   geographically separated.  One may construct the  corresponding  sets of operators
$Z_{i}$, $X_{i}$, $Y_{i}$, as is depicted in  Fig. \ref{Fig:system}(a). We now would like to switch to another partitioning
(e.g., the one depicted in  Fig. \ref{Fig:system}(b)), involving $\tilde{R}$  and $\tilde{L}$ respectively. Trying to express the spin operators associated with this partitioning in terms of the {\it real} spin operators,  we have $\hat{Z}_{\tilde{R}}=\hat{Z}_{L}$, and $\hat{Z}_{\tilde{L}}=\hat{Z}_{L}\otimes \hat{Z}_{R}$. It then follows that 
$Z_{\tilde{R}}$ and $Z_{\tilde{L}}$ cannot be measured by two spatially
separated detectors. This is in stark difference with the foregoing Majorana-based  picture. 

The statement that any state of the system violates the CHSH inequalities in at least one of the
partitionings of Fig. \ref{Fig:system} implies  {\it finite} violation of CHSH inequalities. 
This is quantified by introducing the maximal value of CHSH correlations over the partitioning in  Fig. \ref{Fig:system},  $\mathcal{C}_{0}(\vert\psi\rangle)\equiv\max\left\{ \mathcal{C}_{135|246},\,\mathcal{C}_{564|132},\,\mathcal{C}_{421|563},\,\mathcal{C}_{641|352}\right\}$.
For any $\vert \psi \rangle$,  $\mathcal{C}_{0}(\vert\psi\rangle)-2$ is a positive finite quantity. There is therefore a minimum violation of the CHSH inequality over all states.
From a standard minimization procedure over the parameters $A$, $B$, $C$, $D$ \cite{supplementary}, we obtain
\begin{equation}
\min_{\vert\psi\rangle}\left\{ \mathcal{C}_{0}(\vert\psi\rangle)\right\} \approx 2.031. \label{eq:minimo}
\end{equation}
Note that, since we restrict the analysis here to the four configurations
of Fig. \ref{Fig:system}, the minimum value obtained is in fact a
lower bound of the optimal minimum entanglement.%

\paragraph{Measurement}

\begin{figure}
\includegraphics[width=0.85\columnwidth]{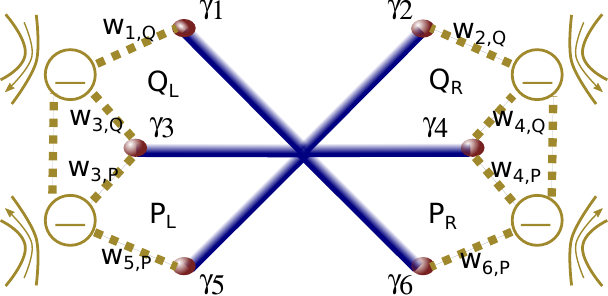}
\caption{Measurement of CHSH correlations in a Multi-terminal Majorana junction. 
The left and right measurement apparatus consist of quantum dots ($Q_L$, $P_L$, $Q_R$, $P_R$) properly coupled to 
the MZM via tunnel coupling (dotted lines) of strength $w_{\alpha,j}$. The charge configuration of each dot is detected by a nearby charge sensor,
schematically depicted as a quantum point contact. The measurement is performed by controlled time pulsed activations of the tunnel coupling $w_{\alpha,j}$.   
\label{Fig:measure}}
\end{figure}

In order to implement the above ideas we need to measure operators of the form  (\ref{eq:operator}) and correlations thereof. While the emerging picture is  quite general, we will demonstrate it by resorting to a specific measurement protocol:  weakly tunnel-coupling quantum dots (QDs) to the multi-terminal Majorana
junction \cite{Flensberg2011,Leijnse2011}, and then measuring their charge. 
Let us describe the measurement procedure for operators associated with 
the $L$ and $R$ Majorana sets. We correspondingly define  $L$- and $R$- detectors,
each consisting of a double quantum dot tunnel-coupled to the three  MZM
in the set, as shown in Fig.\ref{Fig:measure}.
The coupling of the $L$ detector to the corresponding MZM is given  by
the Hamiltonian
\begin{eqnarray}
& H_{{\rm det,L}} =  w_{3,Q}\gamma_{3}(c_{Q,L}-c_{Q,L}^{\dagger})+w_{3,P}\gamma_{3}(c_{P,L}-c_{P,L}^{\dagger}) \nonumber\\
& + w_{1,Q}\gamma_{1}(c_{Q,L}-c_{Q,L}^{\dagger})+w_{5,P}\gamma_{5}(c_{P,L}-c_{P,L}^{\dagger}),
\end{eqnarray}
where $c_{j,L}$, $j=Q,P$ are the electron destruction operators of each  dot of the pair (all electrons are spin polarized) and $w_{\alpha,j}$ 
are the tunneling matrix elements between the superconductor's end-points and the quantum dots. 
 These dots are  tuned such that only one orbital level
per dot is relevant at the energy scales considered. The charge
configuration of the double QD, $(n_{Q,L},n_{P,L}),$ with $n_{j}=0,1$
can be detected by fast charge sensors, e.g. quantum point contacts \cite{Field1993,Elzerman2003,Elzerman2004,Petta2004,Petta2005,Oxtoby2006,Shi2013,Ward2016}. The possibly time dependent tunnel coupling is controlled, e.g. by
a nearby gate voltage \cite{footnote2}.  One initially prepares the decoupled double QD in a generic superposition of singly occupied levels, $\vert \phi_0 \rangle =p_L \vert 0,1 \rangle + q_L \vert 1,0 \rangle $, where $|p_L|^2+|q_L|^2=1$. The tunnel coupling is then switched on for a finite
time $\Delta t$, and is subsequently switched off.  The state of the QDs may be modified, and  the new charge configuration is read out  by the charge sensors. Specifically we access the probability, $\mathcal{P}^L_{\rm (1,0)}$, of finding the double dots in the configuration $(1,0),$
%In principle, the  tunnel coupling may  modify the parity of the Majorana system. 
%To avoid this complication we tune the capacitances of the QD-detector such that only two possible
%detector's readout states are  $(1,0)$ and $(0,1)$.

While the strength and the duration of the QDs-Majorana coupling is adjustable, we consider here, for simplicity, the {\it weak measurement} limit.  
(Going beyond this limit is discussed below \cite{supplementary}.) 
Expanding the time evolution, $U=e^{-iH_{{\rm det,L}}\Delta t}$ , due to the system-detector coupling for small $\Delta t$, and setting for simplicity the initial state of the double QD to $p_L=-iq_L=-i/\sqrt{2}$ and $w_{\alpha,j} \in \mathbb{R}$, the measured probability reads
$\mathcal{P}^L_{{\rm (1,0)}} \approx 1/2 -  \left( \eta_L - \lambda_{L} \langle \hat{O}_{L} \rangle \right) (\Delta t)^2 $
to leading order in $\eta_L = \left (|w_{1,Q}|^2 + |w_{3,Q}|^2 + |w_{3,P}|^2 + |w_{5,P}|^2 \right)/2 $, $\lambda_{L} =[ (w_{1,Q}w_{3,P})^{2} + (w_{3,Q}w_{5,P})^{2} +(w_{1,Q}w_{5,P} )^{2}]^{1/2}$. Here $\hat{O}_{L}$ takes the form of Eq. (\ref{eq:operator}) with $\cos\theta_{L}= w_{1,Q}w_{3,P} /\lambda_L$,
$\sin\theta_{L}\cos\phi_{L}=-w_{3,Q}w_{5,P}/\lambda_L$,
$\sin\theta_{L}\sin\phi_{L}=w_{5,P}w_{1,Q} /\lambda_L$.
For  $\lambda_L \left(\Delta t\right)^{2} \ll 1$ and $\eta_L \left(\Delta t\right)^{2} \ll 1$, this procedure constitutes a weak measurement of the operator $\hat{O}_{L}$. Tuning the parameters $w_{i,j}, i=1,3,5$ and $j=Q,P$  covers all operators of the local algebra of the left ($L$) set.  The same may be repeated to measure the observables represented by the operators $\hat{O}_{R}$ of the right set, and the correlated measurements
implied  by the CHSH inequality are therefore  executable. Specifically, referring
to Fig. \ref{Fig:measure}, one begins with the configuration $p_L=p_R=-i q_L=-i q_R=-i/\sqrt{2}$.
Tunnel coupling the MZM to the QDs, and then  sensing their final configuration,
the probability to end-up in the $(n_{Q,L}=1,n_{P,L}=0,n_{Q,L}=1,n_{Q,R}=0)$ is
\begin{eqnarray}
& \mathcal{P}_{(1,0,1,0)} =  \mathcal{P}^L_{(1,0)}  \mathcal{P}^R_{(1,0)} + \frac{(\Delta t)^4}{6} \left( 
\eta_L^2+\eta_R^2 +\lambda_L^2 + \lambda_R^2 \right. \nonumber \\
 & +2  \left.\eta_L \lambda_L \langle \hat{O}_L\rangle + 2 \eta_R \lambda_R\langle \hat{O}_R \rangle + \lambda_L \lambda_R \langle \hat{O}_L \hat{O}_R\rangle 
\right),
\label{eq:misura-finale}
\end{eqnarray}
where $\lambda_{R}=(w_{2,Q}w_{6,P})^{2} + (w_{2,Q}w_{4,P})^{2} + (w_{4,Q} w_{6,P})^{2} $ and $\eta_R=\left (|w_{2,Q}|^2 + |w_{4,Q}|^2 + |w_{4,P}|^2 + |w_{6,P}|^2 \right)/2$. Eq. \eqref{eq:misura-finale} provides us with a way to evaluate the correlators of the type $\langle \hat{O}_L \hat{O}_R\rangle$ (cf. Eq. \eqref{eq:CHSH}).
To demonstrate how entanglement (violation of CHSH) is detected concretely in our measurement scheme, consider specifically the case of a state in Eq. \eqref{eq:state} prepared with $A=D=0$. The state is maximally entangled in the configuration of Fig. \ref{Fig:system}(a), and the operators $\hat{O}_L$, $\hat{O} '_L$, $\hat{O}_R$, and  $\hat{O}'_R$ required for the maximal violation of the CHSH inequality \cite{Popescu1992,supplementary} are obtained in our scheme by tuning $w_{5,P}=0$, $w_{1,Q}=0$, $w_{2,Q}=-w_{6,P} \ll w_{4,Q}=w_{4,P}$, and $w_{2,Q}=w_{6,P} \ll w_{4,Q}=w_{4,P}$ respectively. The same state, CHSH-tested with the configuration of Fig. \ref{Fig:system}(b), leads to no violation of
the CHSH inequalities. 

A few aspects of the measurement procedure are noteworthy. 
First, since the operators of the left and of the right set
commute, they can be measured simultaneously; non-universal  details of the
time sequence concerning on-and-off switching of the tunneling matrix elements are immaterial. Second, the weak limit of the measurement offers a simple  interpretation
of the results, however the essence of the analysis remains unchanged 
at stronger system-detector interaction, although the calibration of
the detector may become more involved.
Finally, the proposed measurement protocol
requires control of the individual tunnel
matrix elements between the dots and the wires in order to measure
different operators. This  might present an experimental challenge. The required protocol operations may be realized by
variants better suited to experimental implementation, e.g. by controlling individual QD's energy levels, or possibly by Majorana-to-charge conversion measurements \cite{aasen2015}.

\paragraph{Conclusions} We have identified a minimal complexity MZM array: a junction with 6 segments delineating an 8-fold degenerate subspace defined by 6 Majorana zero modes. Unlike paradigmatic quantum states of two spin-$1/2$ particles that may (e.g. spin singlet) or may not (spin triplet-$1$) be entangled, we have shown that {\it any} state in the 4-dimensional fixed-parity degenerate space (e.g., odd parity) is {\it non-locally entangled}. This comes with a minimal bound on the violation of the CHSH inequality (Eq. \eqref{eq:minimo}). We also presented a detector design based on MZM-QD tunnel coupling, amenable to experimental implementation, and showed how to obtain CHSH correlation functions.  Specifically, we discussed the limit of a weak measurement protocol. This ubiquitous non-locality, expressed through non-local entanglement, reflects the intrinsic property of MZM as carriers of fractionalized fermionic degree of freedom. Verification of non-local entanglement requires repeated measurements of CHSH correlations on replica of the same state, with at least 4 different partitioning (into $L$ and $R$ sets)  of the Majorana degrees of freedom. The CHSH inequality will be broken for at least one of these partitioning. 
 
We  acknowledge support from ISF grant 1349/14, DFG grant RO 2247/8-1, CRC 183 of the DFG, the IMOS Israel-Russia program, and the Minerva Foundation.

%\bibliographystyle{nourl_apsrev}
%\bibliography{tqi-Majorana-measurement}

\begin{thebibliography}{42}
\expandafter\ifx\csname natexlab\endcsname\relax\def\natexlab#1{#1}\fi
\expandafter\ifx\csname bibnamefont\endcsname\relax
  \def\bibnamefont#1{#1}\fi
\expandafter\ifx\csname bibfnamefont\endcsname\relax
  \def\bibfnamefont#1{#1}\fi
\expandafter\ifx\csname citenamefont\endcsname\relax
  \def\citenamefont#1{#1}\fi
\providecommand{\bibinfo}[2]{#2}

\bibitem[{\citenamefont{Moore and Read}(1991)}]{Moore1991}
\bibinfo{author}{\bibfnamefont{G.}~\bibnamefont{Moore}} \bibnamefont{and}
  \bibinfo{author}{\bibfnamefont{N.}~\bibnamefont{Read}},
  \bibinfo{journal}{Nucl. Phys. B} \textbf{\bibinfo{volume}{360}},
  \bibinfo{pages}{362} (\bibinfo{year}{1991}).

\bibitem[{\citenamefont{Read and Green}(2000)}]{Read1999}
\bibinfo{author}{\bibfnamefont{N.}~\bibnamefont{Read}} \bibnamefont{and}
  \bibinfo{author}{\bibfnamefont{D.}~\bibnamefont{Green}},
  \bibinfo{journal}{Phys. Rev. B} \textbf{\bibinfo{volume}{61}},
  \bibinfo{pages}{10267} (\bibinfo{year}{2000}).

\bibitem[{\citenamefont{Kitaev}(2001)}]{Kitaev2001}
\bibinfo{author}{\bibfnamefont{A.}~\bibnamefont{Kitaev}},
  \bibinfo{journal}{Physics-Uspekhi} \textbf{\bibinfo{volume}{44}},
  \bibinfo{pages}{131\textbf{}} (\bibinfo{year}{2001}).

\bibitem[{\citenamefont{Lutchyn et~al.}(2010)\citenamefont{Lutchyn, Sau, and
  {Das Sarma}}}]{Lutchyn2010}
\bibinfo{author}{\bibfnamefont{R.~M.} \bibnamefont{Lutchyn}},
  \bibinfo{author}{\bibfnamefont{J.~D.} \bibnamefont{Sau}}, \bibnamefont{and}
  \bibinfo{author}{\bibfnamefont{S.}~\bibnamefont{{Das Sarma}}},
  \bibinfo{journal}{Phys. Rev. Lett.} \textbf{\bibinfo{volume}{105}},
  \bibinfo{pages}{077001} (\bibinfo{year}{2010}).

\bibitem[{\citenamefont{Oreg et~al.}(2010)\citenamefont{Oreg, Refael, and von
  Oppen}}]{Oreg2010}
\bibinfo{author}{\bibfnamefont{Y.}~\bibnamefont{Oreg}},
  \bibinfo{author}{\bibfnamefont{G.}~\bibnamefont{Refael}}, \bibnamefont{and}
  \bibinfo{author}{\bibfnamefont{F.}~\bibnamefont{von Oppen}},
  \bibinfo{journal}{Phys. Rev. Lett.} \textbf{\bibinfo{volume}{105}},
  \bibinfo{pages}{177002} (\bibinfo{year}{2010}).

\bibitem[{\citenamefont{Wilczek}(2009)}]{Wilczek2009}
\bibinfo{author}{\bibfnamefont{F.}~\bibnamefont{Wilczek}},
  \bibinfo{journal}{Nat. Phys.} \textbf{\bibinfo{volume}{5}},
  \bibinfo{pages}{614} (\bibinfo{year}{2009}).

\bibitem[{\citenamefont{Wilczek}(1982)}]{Wilczek1982}
\bibinfo{author}{\bibfnamefont{F.}~\bibnamefont{Wilczek}},
  \bibinfo{journal}{Phys. Rev. Lett.} \textbf{\bibinfo{volume}{49}},
  \bibinfo{pages}{957} (\bibinfo{year}{1982}).

\bibitem[{\citenamefont{Stern}(2010)}]{Stern2010}
\bibinfo{author}{\bibfnamefont{A.}~\bibnamefont{Stern}},
  \bibinfo{journal}{Nature} \textbf{\bibinfo{volume}{464}},
  \bibinfo{pages}{187} (\bibinfo{year}{2010}).

\bibitem[{\citenamefont{Nayak et~al.}(2008)\citenamefont{Nayak, Simon, Stern,
  Freedman, and {Das Sarma}}}]{Nayak2008}
\bibinfo{author}{\bibfnamefont{C.}~\bibnamefont{Nayak}},
  \bibinfo{author}{\bibfnamefont{S.~H.} \bibnamefont{Simon}},
  \bibinfo{author}{\bibfnamefont{A.}~\bibnamefont{Stern}},
  \bibinfo{author}{\bibfnamefont{M.}~\bibnamefont{Freedman}}, \bibnamefont{and}
  \bibinfo{author}{\bibfnamefont{S.}~\bibnamefont{{Das Sarma}}},
  \bibinfo{journal}{Rev. Mod. Phys.} \textbf{\bibinfo{volume}{80}},
  \bibinfo{pages}{1083} (\bibinfo{year}{2008}).

\bibitem[{\citenamefont{Mourik et~al.}(2012)\citenamefont{Mourik, Zuo, Frolov,
  Plissard, Bakkers, and Kouwenhoven}}]{Mourik2012}
\bibinfo{author}{\bibfnamefont{V.}~\bibnamefont{Mourik}},
  \bibinfo{author}{\bibfnamefont{K.}~\bibnamefont{Zuo}},
  \bibinfo{author}{\bibfnamefont{S.~M.} \bibnamefont{Frolov}},
  \bibinfo{author}{\bibfnamefont{S.~R.} \bibnamefont{Plissard}},
  \bibinfo{author}{\bibfnamefont{E.~P. A.~M.} \bibnamefont{Bakkers}},
  \bibnamefont{and} \bibinfo{author}{\bibfnamefont{L.~P.}
  \bibnamefont{Kouwenhoven}}, \bibinfo{journal}{Science}
  \textbf{\bibinfo{volume}{336}}, \bibinfo{pages}{1003} (\bibinfo{year}{2012}).

\bibitem[{\citenamefont{Das et~al.}(2012)\citenamefont{Das, Ronen, Most, Oreg,
  Heiblum, and Shtrikman}}]{Das2012}
\bibinfo{author}{\bibfnamefont{A.}~\bibnamefont{Das}},
  \bibinfo{author}{\bibfnamefont{Y.}~\bibnamefont{Ronen}},
  \bibinfo{author}{\bibfnamefont{Y.}~\bibnamefont{Most}},
  \bibinfo{author}{\bibfnamefont{Y.}~\bibnamefont{Oreg}},
  \bibinfo{author}{\bibfnamefont{M.}~\bibnamefont{Heiblum}}, \bibnamefont{and}
  \bibinfo{author}{\bibfnamefont{H.}~\bibnamefont{Shtrikman}},
  \bibinfo{journal}{Nat. Phys.} \textbf{\bibinfo{volume}{8}},
  \bibinfo{pages}{887} (\bibinfo{year}{2012}).

\bibitem[{\citenamefont{Churchill et~al.}(2013)\citenamefont{Churchill, Fatemi,
  Grove-Rasmussen, Deng, Caroff, Xu, and Marcus}}]{Churchill2013}
\bibinfo{author}{\bibfnamefont{H.~O.~H.} \bibnamefont{Churchill}},
  \bibinfo{author}{\bibfnamefont{V.}~\bibnamefont{Fatemi}},
  \bibinfo{author}{\bibfnamefont{K.}~\bibnamefont{Grove-Rasmussen}},
  \bibinfo{author}{\bibfnamefont{M.~T.} \bibnamefont{Deng}},
  \bibinfo{author}{\bibfnamefont{P.}~\bibnamefont{Caroff}},
  \bibinfo{author}{\bibfnamefont{H.~Q.} \bibnamefont{Xu}}, \bibnamefont{and}
  \bibinfo{author}{\bibfnamefont{C.~M.} \bibnamefont{Marcus}},
  \bibinfo{journal}{Phys. Rev. B}
  \textbf{\bibinfo{volume}{87}}, \bibinfo{pages}{241401}
  (\bibinfo{year}{2013}).

\bibitem[{\citenamefont{Nadj-Perge et~al.}(2014)\citenamefont{Nadj-Perge,
  Drozdov, Li, Chen, Jeon, Seo, MacDonald, Bernevig, and
  Yazdani}}]{Nadj-Perge2014}
\bibinfo{author}{\bibfnamefont{S.}~\bibnamefont{Nadj-Perge}},
  \bibinfo{author}{\bibfnamefont{I.~K.} \bibnamefont{Drozdov}},
  \bibinfo{author}{\bibfnamefont{J.}~\bibnamefont{Li}},
  \bibinfo{author}{\bibfnamefont{H.}~\bibnamefont{Chen}},
  \bibinfo{author}{\bibfnamefont{S.}~\bibnamefont{Jeon}},
  \bibinfo{author}{\bibfnamefont{J.}~\bibnamefont{Seo}},
  \bibinfo{author}{\bibfnamefont{A.~H.} \bibnamefont{MacDonald}},
  \bibinfo{author}{\bibfnamefont{B.~A.} \bibnamefont{Bernevig}},
  \bibnamefont{and} \bibinfo{author}{\bibfnamefont{A.}~\bibnamefont{Yazdani}},
  \bibinfo{journal}{Science} \textbf{\bibinfo{volume}{346}},
  \bibinfo{pages}{602} (\bibinfo{year}{2014}).

\bibitem[{\citenamefont{Albrecht et~al.}(2016)\citenamefont{Albrecht,
  Higginbotham, Madsen, Kuemmeth, Jespersen, Nyg{\aa}rd, Krogstrup, and
  Marcus}}]{Albrecht2016}
\bibinfo{author}{\bibfnamefont{S.~M.} \bibnamefont{Albrecht}},
  \bibinfo{author}{\bibfnamefont{A.~P.} \bibnamefont{Higginbotham}},
  \bibinfo{author}{\bibfnamefont{M.}~\bibnamefont{Madsen}},
  \bibinfo{author}{\bibfnamefont{F.}~\bibnamefont{Kuemmeth}},
  \bibinfo{author}{\bibfnamefont{T.~S.} \bibnamefont{Jespersen}},
  \bibinfo{author}{\bibfnamefont{J.}~\bibnamefont{Nyg{\aa}rd}},
  \bibinfo{author}{\bibfnamefont{P.}~\bibnamefont{Krogstrup}},
  \bibnamefont{and} \bibinfo{author}{\bibfnamefont{C.~M.}
  \bibnamefont{Marcus}}, \bibinfo{journal}{Nature}
  \textbf{\bibinfo{volume}{531}}, \bibinfo{pages}{206} (\bibinfo{year}{2016}).
  
  \bibitem{Deng2016}
 M. T. Deng, S. Vaitiek\`enas, E. B. Hansen, J. Danon, M. Leijnse, K. Flensberg, J. Nygard, P. Krogstrup, C. M. Marcus,
Science {\bf 354}, 1557 (2016).

\bibitem[{\citenamefont{Aasen et~al.}(2016)\citenamefont{Aasen, Hell, Mishmash,
  Higginbotham, Danon, Leijnse, Jespersen, Folk, Marcus, Flensberg
  et~al.}}]{aasen2015}
\bibinfo{author}{\bibfnamefont{D.}~\bibnamefont{Aasen}},
  \bibinfo{author}{\bibfnamefont{M.}~\bibnamefont{Hell}},
  \bibinfo{author}{\bibfnamefont{R.~V.} \bibnamefont{Mishmash}},
  \bibinfo{author}{\bibfnamefont{A.}~\bibnamefont{Higginbotham}},
  \bibinfo{author}{\bibfnamefont{J.}~\bibnamefont{Danon}},
  \bibinfo{author}{\bibfnamefont{M.}~\bibnamefont{Leijnse}},
  \bibinfo{author}{\bibfnamefont{T.~S.} \bibnamefont{Jespersen}},
  \bibinfo{author}{\bibfnamefont{J.~A.} \bibnamefont{Folk}},
  \bibinfo{author}{\bibfnamefont{C.~M.} \bibnamefont{Marcus}},
  \bibinfo{author}{\bibfnamefont{K.}~\bibnamefont{Flensberg}},
  \bibnamefont{et~al.}, \bibinfo{journal}{Phys. Rev. X}
  \textbf{\bibinfo{volume}{6}}, \bibinfo{pages}{031016} (\bibinfo{year}{2016}).


\bibitem{Plugge2016}
S. Plugge, L. A. Landau, E. Sela, A. Altland, K. Flensberg, and R. Egger,  Physical Review B,{\bf 94}, 174514 (2016).

%\bibitem[{\citenamefont{Plugge et~al.}(2016{\natexlab{a}})\citenamefont{Plugge,
%  Landau, Sela, Altland, Flensberg, and Egger}}]{Plugge2016}
%\bibinfo{author}{\bibfnamefont{S.}~\bibnamefont{Plugge}},
%  \bibinfo{author}{\bibfnamefont{L.~A.} \bibnamefont{Landau}},
%  \bibinfo{author}{\bibfnamefont{E.}~\bibnamefont{Sela}},
%  \bibinfo{author}{\bibfnamefont{A.}~\bibnamefont{Altland}},
%  \bibinfo{author}{\bibfnamefont{K.}~\bibnamefont{Flensberg}},
%  \bibnamefont{and} \bibinfo{author}{\bibfnamefont{R.}~\bibnamefont{Egger}},
%  pp. \bibinfo{pages}{1--23} (\bibinfo{year}{2016}{\natexlab{a}}).

\bibitem{Plugge2016a}
 S. Plugge, A. Rasmussen, R. Egger, and K. Flensberg, New J. Phys. {\bf 19}, 012001 (2017).

%\bibitem[{\citenamefont{Plugge et~al.}(2016{\natexlab{b}})\citenamefont{Plugge,
%  Rasmussen, Egger, and Flensberg}}]{Plugge2016a}
%\bibinfo{author}{\bibfnamefont{S.}~\bibnamefont{Plugge}},
%  \bibinfo{author}{\bibfnamefont{A.}~\bibnamefont{Rasmussen}},
%  \bibinfo{author}{\bibfnamefont{R.}~\bibnamefont{Egger}}, \bibnamefont{and}
%  \bibinfo{author}{\bibfnamefont{K.}~\bibnamefont{Flensberg}}, pp.
%  \bibinfo{pages}{1--10} (\bibinfo{year}{2016}{\natexlab{b}}),
%  \eprint{1609.01697}.

\bibitem[{\citenamefont{Fu}(2010)}]{Fu2010}
\bibinfo{author}{\bibfnamefont{L.}~\bibnamefont{Fu}}, \bibinfo{journal}{Phys.
  Rev. Lett.} \textbf{\bibinfo{volume}{104}}, \bibinfo{pages}{056402}
  (\bibinfo{year}{2010}).


\bibitem[{\citenamefont{Michaeli et~al.}(2016)\citenamefont{Michaeli, Landau,
  Sela, and Fu}}]{Michaeli2016}
\bibinfo{author}{\bibfnamefont{K.}~\bibnamefont{Michaeli}},
  \bibinfo{author}{\bibfnamefont{L.~A.} \bibnamefont{Landau}},
  \bibinfo{author}{\bibfnamefont{E.}~\bibnamefont{Sela}}, \bibnamefont{and}
  \bibinfo{author}{\bibfnamefont{L.}~\bibnamefont{Fu}}, arXiv:\eprint{1608.00581} (2016).

\bibitem{Vijay2016}
S. Vijay, and L. Fu, Physical Review B {\bf 94}, 235446 (2016).

%\bibitem[{\citenamefont{Vijay and Fu}(2016)}]{Vijay2016}
%\bibinfo{author}{\bibfnamefont{S.}~\bibnamefont{Vijay}} \bibnamefont{and}
%  \bibinfo{author}{\bibfnamefont{L.}~\bibnamefont{Fu}} (\bibinfo{year}{2016}),
%  \eprint{1609.00950}.

\bibitem[{\citenamefont{Zocher and Rosenow}(2013)}]{Rosenow2012}
\bibinfo{author}{\bibfnamefont{B.}~\bibnamefont{Zocher}}, \bibnamefont{and}
  \bibinfo{author}{\bibfnamefont{B.}~\bibnamefont{Rosenow}},
  \bibinfo{journal}{Phys. Rev. Lett.} \textbf{\bibinfo{volume}{111}},
  \bibinfo{pages}{036802} (\bibinfo{year}{2013}).

\bibitem[{\citenamefont{Rubbert and Akhmerov}(2016)}]{Rubbert2016}
\bibinfo{author}{\bibfnamefont{S.}~\bibnamefont{Rubbert}} \bibnamefont{and}
  \bibinfo{author}{\bibfnamefont{A.~R.} \bibnamefont{Akhmerov}},
  \bibinfo{journal}{Phys. Rev. B} \textbf{\bibinfo{volume}{94}},
  \bibinfo{pages}{115430} (\bibinfo{year}{2016}).

%\bibitem[{Note1()}]{Note1}
%Note1, \bibinfo{note}{hereafter we refer to entanglement of spatially distinct
%  degrees of freedom.}
\bibitem{footnote1}
Hereafter we refer to entanglement of spatially distinct degrees of freedom.

\bibitem[{\citenamefont{Bell}(1965)}]{Bell1965}
\bibinfo{author}{\bibfnamefont{J.}~\bibnamefont{Bell}},
  \bibinfo{journal}{Physics} \textbf{\bibinfo{volume}{1}},
  \bibinfo{pages}{195} (\bibinfo{year}{1965}).

\bibitem[{\citenamefont{Giustina et~al.}(2015)\citenamefont{Giustina,
  Versteegh, Wengerowsky, Handsteiner, Hochrainer, Phelan, Steinlechner,
  Kofler, Larsson, Abell{\'{a}}n et~al.}}]{Giustina2015}
\bibinfo{author}{\bibfnamefont{M.}~\bibnamefont{Giustina}},
  \bibinfo{author}{\bibfnamefont{M.~A.~M.} \bibnamefont{Versteegh}},
  \bibinfo{author}{\bibfnamefont{S.}~\bibnamefont{Wengerowsky}},
  \bibinfo{author}{\bibfnamefont{J.}~\bibnamefont{Handsteiner}},
  \bibinfo{author}{\bibfnamefont{A.}~\bibnamefont{Hochrainer}},
  \bibinfo{author}{\bibfnamefont{K.}~\bibnamefont{Phelan}},
  \bibinfo{author}{\bibfnamefont{F.}~\bibnamefont{Steinlechner}},
  \bibinfo{author}{\bibfnamefont{J.}~\bibnamefont{Kofler}},
  \bibinfo{author}{\bibfnamefont{J.-{\AA}.} \bibnamefont{Larsson}},
  \bibinfo{author}{\bibfnamefont{C.}~\bibnamefont{Abell{\'{a}}n}},
  \bibnamefont{et~al.}, \bibinfo{journal}{Phys. Rev. Lett.}
  \textbf{\bibinfo{volume}{115}}, \bibinfo{pages}{250401}
  (\bibinfo{year}{2015}).

\bibitem[{\citenamefont{Hensen et~al.}(2015)\citenamefont{Hensen, Bernien,
  Dr{\'{e}}au, Reiserer, Kalb, Blok, Ruitenberg, Vermeulen, Schouten,
  Abell{\'{a}}n et~al.}}]{Hensen2015}
\bibinfo{author}{\bibfnamefont{B.}~\bibnamefont{Hensen}},
  \bibinfo{author}{\bibfnamefont{H.}~\bibnamefont{Bernien}},
  \bibinfo{author}{\bibfnamefont{A.~E.} \bibnamefont{Dr{\'{e}}au}},
  \bibinfo{author}{\bibfnamefont{A.}~\bibnamefont{Reiserer}},
  \bibinfo{author}{\bibfnamefont{N.}~\bibnamefont{Kalb}},
  \bibinfo{author}{\bibfnamefont{M.~S.} \bibnamefont{Blok}},
  \bibinfo{author}{\bibfnamefont{J.}~\bibnamefont{Ruitenberg}},
  \bibinfo{author}{\bibfnamefont{R.~F.~L.} \bibnamefont{Vermeulen}},
  \bibinfo{author}{\bibfnamefont{R.~N.} \bibnamefont{Schouten}},
  \bibinfo{author}{\bibfnamefont{C.}~\bibnamefont{Abell{\'{a}}n}},
  \bibnamefont{et~al.}, \bibinfo{journal}{Nature}
  \textbf{\bibinfo{volume}{526}}, \bibinfo{pages}{682} (\bibinfo{year}{2015}).

\bibitem[{\citenamefont{Clauser et~al.}(1969)\citenamefont{Clauser, Horne,
  Shimony, and Holt}}]{Clauser1969}
\bibinfo{author}{\bibfnamefont{J.~F.} \bibnamefont{Clauser}},
  \bibinfo{author}{\bibfnamefont{M.~A.} \bibnamefont{Horne}},
  \bibinfo{author}{\bibfnamefont{A.}~\bibnamefont{Shimony}}, \bibnamefont{and}
  \bibinfo{author}{\bibfnamefont{R.~A.} \bibnamefont{Holt}},
  \bibinfo{journal}{Phys. Rev. Lett.} \textbf{\bibinfo{volume}{23}},
  \bibinfo{pages}{880} (\bibinfo{year}{1969}).
  
  
  \bibitem{Campbell2014}
  E. T. Campbell, M. J. Hoban, J. Eisert, Quant. Inf. Comp. {\bf 14}, 0981 (2014).
  
  \bibitem{Deng2013}
  D. Deng and L. Duan, Phys. Rev. A {\bf 88}, 012323 (2013)
   \bibitem[{\citenamefont{Clarke et~al.}(2016)\citenamefont{Clarke, Sau, and {Das
  			Sarma}}}]{Clarke2015}
  \bibinfo{author}{\bibfnamefont{D.~J.} \bibnamefont{Clarke}},
  \bibinfo{author}{\bibfnamefont{J.~D.} \bibnamefont{Sau}}, \bibnamefont{and}
  \bibinfo{author}{\bibfnamefont{S.}~\bibnamefont{{Das Sarma}}},
  \bibinfo{journal}{Phys. Rev. X} \textbf{\bibinfo{volume}{6}},
  \bibinfo{pages}{021005} (\bibinfo{year}{2016}).
  
 

\bibitem[{\citenamefont{Alicea et~al.}(2011)\citenamefont{Alicea, Oreg, Refael,
  von Oppen, and Fisher}}]{Alicea2011}
\bibinfo{author}{\bibfnamefont{J.}~\bibnamefont{Alicea}},
  \bibinfo{author}{\bibfnamefont{Y.}~\bibnamefont{Oreg}},
  \bibinfo{author}{\bibfnamefont{G.}~\bibnamefont{Refael}},
  \bibinfo{author}{\bibfnamefont{F.}~\bibnamefont{von Oppen}},
  \bibnamefont{and} \bibinfo{author}{\bibfnamefont{M.~P.~a.}
  \bibnamefont{Fisher}}, \bibinfo{journal}{Nat. Phys.}
  \textbf{\bibinfo{volume}{7}}, \bibinfo{pages}{412} (\bibinfo{year}{2011}).

\bibitem[{\citenamefont{Cirel'son}(1980)}]{Cirelson1980}
\bibinfo{author}{\bibfnamefont{B.~S.} \bibnamefont{Cirel'son}},
  \bibinfo{journal}{Lett. Math. Phys.} \textbf{\bibinfo{volume}{4}},
  \bibinfo{pages}{93} (\bibinfo{year}{1980}).
  
  \bibitem{supplementary}
  See supplementary material.

\bibitem[{\citenamefont{Flensberg}(2011)}]{Flensberg2011}
\bibinfo{author}{\bibfnamefont{K.}~\bibnamefont{Flensberg}},
  \bibinfo{journal}{Phys. Rev. Lett.} \textbf{\bibinfo{volume}{106}},
  \bibinfo{pages}{090503} (\bibinfo{year}{2011}).

\bibitem[{\citenamefont{Leijnse and Flensberg}(2011)}]{Leijnse2011}
\bibinfo{author}{\bibfnamefont{M.}~\bibnamefont{Leijnse}} \bibnamefont{and}
  \bibinfo{author}{\bibfnamefont{K.}~\bibnamefont{Flensberg}},
  \bibinfo{journal}{Phys. Rev. Lett.} \textbf{\bibinfo{volume}{107}},
  \bibinfo{pages}{210502} (\bibinfo{year}{2011}).

\bibitem[{\citenamefont{Field et~al.}(1993)\citenamefont{Field, Smith, Pepper,
  Ritchie, Frost, Jones, and Hasko}}]{Field1993}
\bibinfo{author}{\bibfnamefont{M.}~\bibnamefont{Field}},
  \bibinfo{author}{\bibfnamefont{C. G.}~\bibnamefont{Smith}},
  \bibinfo{author}{\bibfnamefont{M.}~\bibnamefont{Pepper}},
  \bibinfo{author}{\bibfnamefont{D. A.}~\bibnamefont{Ritchie}},
  \bibinfo{author}{\bibfnamefont{J. E. F.}~\bibnamefont{Frost}},
  \bibinfo{author}{\bibfnamefont{G. A. C.}~\bibnamefont{Jones}}, \bibnamefont{and}
  \bibinfo{author}{\bibfnamefont{D. G.}~\bibnamefont{Hasko}},
  \bibinfo{journal}{Phys. Rev. Lett.} \textbf{\bibinfo{volume}{70}},
  \bibinfo{pages}{1311} (\bibinfo{year}{1993}).

\bibitem[{\citenamefont{Elzerman et~al.}(2003)\citenamefont{Elzerman, Hanson,
  Greidanus, {Willems van Beveren}, {De Franceschi}, Vandersypen, Tarucha, and
  Kouwenhoven}}]{Elzerman2003}
\bibinfo{author}{\bibfnamefont{J.~M.} \bibnamefont{Elzerman}},
  \bibinfo{author}{\bibfnamefont{R.}~\bibnamefont{Hanson}},
  \bibinfo{author}{\bibfnamefont{J.~S.} \bibnamefont{Greidanus}},
  \bibinfo{author}{\bibfnamefont{L.~H.~Willems} \bibnamefont{{van Beveren}}},
  \bibinfo{author}{\bibfnamefont{S.}~\bibnamefont{{De Franceschi}}},
  \bibinfo{author}{\bibfnamefont{L.~M.~K.} \bibnamefont{Vandersypen}},
  \bibinfo{author}{\bibfnamefont{S.}~\bibnamefont{Tarucha}}, \bibnamefont{and}
  \bibinfo{author}{\bibfnamefont{L.~P.} \bibnamefont{Kouwenhoven}},
  \bibinfo{journal}{Phys. Rev. B} \textbf{\bibinfo{volume}{67}},
  \bibinfo{pages}{161308} (\bibinfo{year}{2003}).

\bibitem[{\citenamefont{Elzerman et~al.}(2004)\citenamefont{Elzerman, Hanson,
  {Willems van Beveren}, Witkamp, Vandersypen, and Kouwenhoven}}]{Elzerman2004}
\bibinfo{author}{\bibfnamefont{J.~M.} \bibnamefont{Elzerman}},
  \bibinfo{author}{\bibfnamefont{R.}~\bibnamefont{Hanson}},
  \bibinfo{author}{\bibfnamefont{L.~H.} \bibnamefont{{Willems van Beveren}}},
  \bibinfo{author}{\bibfnamefont{B.}~\bibnamefont{Witkamp}},
  \bibinfo{author}{\bibfnamefont{L.~M.~K.} \bibnamefont{Vandersypen}},
  \bibnamefont{and} \bibinfo{author}{\bibfnamefont{L.~P.}
  \bibnamefont{Kouwenhoven}}, \bibinfo{journal}{Nature}
  \textbf{\bibinfo{volume}{430}}, \bibinfo{pages}{431} (\bibinfo{year}{2004}).

\bibitem[{\citenamefont{Petta et~al.}(2004)\citenamefont{Petta, Johnson,
  Marcus, Hanson, and Gossard}}]{Petta2004}
\bibinfo{author}{\bibfnamefont{J.~R.} \bibnamefont{Petta}},
  \bibinfo{author}{\bibfnamefont{A.~C.} \bibnamefont{Johnson}},
  \bibinfo{author}{\bibfnamefont{C.~M.} \bibnamefont{Marcus}},
  \bibinfo{author}{\bibfnamefont{M.~P.} \bibnamefont{Hanson}},
  \bibnamefont{and} \bibinfo{author}{\bibfnamefont{A.~C.}
  \bibnamefont{Gossard}}, \bibinfo{journal}{Phys. Rev. Lett.}
  \textbf{\bibinfo{volume}{93}}, \bibinfo{pages}{186802}
  (\bibinfo{year}{2004}).

\bibitem[{\citenamefont{Petta}(2005)}]{Petta2005}
\bibinfo{author}{\bibfnamefont{J.~R.} \bibnamefont{Petta}},
  \bibinfo{journal}{Science} \textbf{\bibinfo{volume}{309}},
  \bibinfo{pages}{2180} (\bibinfo{year}{2005}).

\bibitem[{\citenamefont{Oxtoby et~al.}(2006)\citenamefont{Oxtoby, Wiseman, and
  Sun}}]{Oxtoby2006}
\bibinfo{author}{\bibfnamefont{N.~P.} \bibnamefont{Oxtoby}},
  \bibinfo{author}{\bibfnamefont{H.~M.} \bibnamefont{Wiseman}},
  \bibnamefont{and} \bibinfo{author}{\bibfnamefont{H.-B.} \bibnamefont{Sun}},
  \bibinfo{journal}{Phys. Rev. B} \textbf{\bibinfo{volume}{74}},
  \bibinfo{pages}{045328} (\bibinfo{year}{2006}).

\bibitem[{\citenamefont{Shi et~al.}(2013)\citenamefont{Shi, Simmons, Ward,
  Prance, Mohr, Koh, Gamble, Wu, Savage, Lagally et~al.}}]{Shi2013}
\bibinfo{author}{\bibfnamefont{Z.}~\bibnamefont{Shi}},
  \bibinfo{author}{\bibfnamefont{C.~B.} \bibnamefont{Simmons}},
  \bibinfo{author}{\bibfnamefont{D.~R.} \bibnamefont{Ward}},
  \bibinfo{author}{\bibfnamefont{J.~R.} \bibnamefont{Prance}},
  \bibinfo{author}{\bibfnamefont{R.~T.} \bibnamefont{Mohr}},
  \bibinfo{author}{\bibfnamefont{T.~S.} \bibnamefont{Koh}},
  \bibinfo{author}{\bibfnamefont{J.~K.} \bibnamefont{Gamble}},
  \bibinfo{author}{\bibfnamefont{X.}~\bibnamefont{Wu}},
  \bibinfo{author}{\bibfnamefont{D.~E.} \bibnamefont{Savage}},
  \bibinfo{author}{\bibfnamefont{M.~G.} \bibnamefont{Lagally}},
  \bibnamefont{et~al.}, \bibinfo{journal}{Phys. Rev. B}
  \textbf{\bibinfo{volume}{88}}, \bibinfo{pages}{075416}
  (\bibinfo{year}{2013}).

\bibitem[{\citenamefont{Ward et~al.}(2016)\citenamefont{Ward, Kim, Savage,
  Lagally, Foote, Friesen, Coppersmith, and Eriksson}}]{Ward2016}
\bibinfo{author}{\bibfnamefont{D.~R.} \bibnamefont{Ward}},
  \bibinfo{author}{\bibfnamefont{D.}~\bibnamefont{Kim}},
  \bibinfo{author}{\bibfnamefont{D.~E.} \bibnamefont{Savage}},
  \bibinfo{author}{\bibfnamefont{M.~G.} \bibnamefont{Lagally}},
  \bibinfo{author}{\bibfnamefont{H.}~\bibnamefont{Foote}},
  \bibinfo{author}{\bibfnamefont{M.}~\bibnamefont{Friesen}},
  \bibinfo{author}{\bibfnamefont{S.~N.} \bibnamefont{Coppersmith}},
  \bibnamefont{and} \bibinfo{author}{\bibfnamefont{M.~A.}
  \bibnamefont{Eriksson}}, \bibinfo{journal}{Npj Quantum Information} 
  \textbf{\bibinfo{volume}{2}}, \bibinfo{pages}{16032} (\bibinfo{year}{2016}).
  
\bibitem{footnote2}
Note that we do not demand here a time-dependent control
of the individual tunnel coupling matrix elements, but rather an overall
on-and-off switching of the QDs-Majorana coupling. The relative strength of  the tunnel couplings
$w_{j,k}$ can be fixed, e.g., at  the fabrication stage.
\bibitem{Popescu1992}
S. Popescu and D. Rohrlich, Physics Letters A, {\bf 166}, 293 (1992).

%\bibitem[{Note2()}]{Note2}
%Note2, \bibinfo{note}{note that we do not demand here a time-dependent control
%  of the individual tunnel coupling matrix elements, but rather an overall, on
%  and off switching of the QDs-Majorana coupling. The relative strength of the
%  tunnel couplings $w_{j,k}$ can be fixed. e.g. at the fabrication stage.}



%\bibitem[{Note3()}]{Note3}
%Note3, \bibinfo{note}{footnote: It has been shown there that braiding
%  operations on a 6 - \st{MBS} \aleadd{MZM} setup will not be capable to produce entanglement in
%  the computational basis, cf. A practical phase gate for producing Bell
%  violations in Majorana wires David J. Clarke, Jay D. Sau, Sankar Das Sarma
%  \cite {Clarke2015}}.


\end{thebibliography}

\newpage

\renewcommand{\thesection}{S\arabic{section}}  
\renewcommand{\thetable}{S\arabic{table}}  
\renewcommand{\thefigure}{S\arabic{figure}} 
\renewcommand{\theequation}{S\arabic{equation}} 

\onecolumngrid
\section*{Supplementary Material}

\section{Measurement protocol beyond the weak limit}

In the manuscript we analyze the measurement protocol in the weak measurement regime. Specifically we consider there the probability for the transition of the detector's dots states from $(i \vert 0,1 \rangle_L + \vert 1,0 \rangle_L)/\sqrt{2} \otimes (i \vert 0,1 \rangle_R + \vert 1,0 \rangle_R)/\sqrt{2}$ to  $ \vert 1,0 \rangle_L \otimes  \vert 1,0 \rangle_R$. 
The transition probability is extracted by detecting the dots configuration $(n_{Q,L},n_{P,L},n_{Q,R},n_{P,R}) = (1,0,1,0)$. The weak measurement regime is obtained for small system-detector coupling and/or short measurement time. The transition probability is then dominated by the leading order term in $(\Delta t)^4$. This regime allows to conveniently relate the measured probabilities to the system's observables [Eq. (8)] entering the CHSH inequalities. Going beyond the weak measurement regime makes the relation to the system's observables more complicated, but it  does not undermine the validity of the measurement protocol. Here we analyze the measurement signal for arbitrary parameters' strengths.

The analysis of the general-strength measurement can be formulated in terms of generalised positive operator valued measurements (POVM), which describe all  possible outcomes of the charge configurations of the dots. The different outcomes of a charge measurement are then labelled by all possible combinations of $n_{j}=0,1$. The process can formally be described in terms of Kraus operators 
\begin{small}
	\begin{gather}
	\label{eq:grande}
	M_{(n_{Q,L},n_{P,L},n_{Q,R},n_{P,R})} = M_{L,(n_{Q,L},n_{P,L})} \cdot M_{R,(n_{Q,R},n_{P,R})}, \\
	M_{L,(n_{Q,L},n_{P,L})} = \tr_{\rm det L} \left\{ \vert \Psi_{L,(n_{Q,L},n_{P,L})} \rangle\langle \Psi_{L,(n_{Q,L},n_{P,L})} \vert \right\}, \\
	M_{R,(n_{Q,R},n_{P,R})}= \tr_{\rm det R} \left\{ \vert \Psi_{R,(n_{Q,R},n_{P,R})} \rangle\langle \Psi_{R,(n_{Q,R},n_{P,R})} \vert \right\} , \\
	\begin{split}
	\vert \Psi_{L,(n_{Q,L},n_{P,L})} \rangle = & \left( c_{Q,L}^\dagger c_{Q,L} \right )^{n_{Q,L}}\left( c_{Q,L} c_{Q,L}^\dagger \right )^{1-n_{Q,L}} \left( c_{P,L}^\dagger c_{P,L} \right )^{n_{P,L}}\left( c_{P,L} c_{P,L}^\dagger \right )^{1-n_{P,L}}  \\ 
	& \cdot \exp \left( -iH_{\rm det,L} \Delta t \right) (q_L c_{Q,L}^\dagger + p_L c_{P,L}^\dagger)  \vert 0\rangle_L , \\
	\vert \Psi_{R,(n_{Q,R},n_{P,R})} \rangle = &\left( c_{Q,R}^\dagger c_{Q,R} \right )^{n_{Q,R}} \left( c_{Q,R} c_{Q,R}^\dagger \right )^{1-n_{q,R}} \left( c_{P,R}^\dagger c_{P,R} \right )^{n_{P,R}}\left( c_{P,R} c_{P,R}^\dagger \right )^{1-n_{P,R}}  \\  & \cdot \exp \left( -iH_{\rm det,R} \Delta t \right) ( q_R c_{Q,R}^\dagger + p_Rc_{P,R}^\dagger)  \vert 0\rangle_R ,
	\end{split}
	\end{gather}
\end{small}
where $\vert 0\rangle_L$, $\vert 0\rangle_L$ are the state of empty left and right dots respectively, and $\tr_{\rm det L}$ $\tr_{\rm det R}$ denotes the trace over the degrees of freedoms of the left and right dots (detectors).
The probability to obtain a specific outcome $(n_{A,L},n_{B,L},n_{A,R},n_{B,R})$  and the corresponding state after the measurement are given by
\begin{gather}
P_{(n_{Q,L},n_{P,L},n_{Q,R},n_{P,R})} = \tr \left\{ M_{(n_{Q,L},n_{P,L},n_{Q,R},n_{P,R})} \rho M^{\dagger}_{(n_{Q,L},n_{P,L},n_{Q,R},n_{P,R})} \right\} , \label{eq:A1}\\
\rho'=M_{(n_{Q,L},n_{P,L},n_{Q,R},n_{P,R})} 
\rho M^{\dagger}_{(n_{Q,L},n_{P,L},n_{Q,R},n_{P,R})}
/P_{(n_{Q,L},n_{P,L},n_{Q,R},n_{P,R})}.
\end{gather}

The signal we are interested in is determined by the Kraus operator $M_{(1,0,1,0)}$. The expression for $M_{(1,0,1,0)}$ can be simplified considerably by noting that $[H_{\rm det,L},H_{\rm det,R}]= 0$, and that the  left and right dots are detected and prepared in states of identical parity. One can then express 
\begin{equation}
\vert \Psi_{L,(1,0)} \rangle = \left( c_{Q,L}^\dagger c_{Q,L} \right ) \sum_{m=0}^\infty \frac{(-\Delta t^2)^m}{(2m)!} H_{\rm det,L}^{2m} (q_L c_{Q,L}^\dagger + p_L c_{P,L}^\dagger)  \vert 0\rangle_L , \label{mela}
\end{equation}
where $H^2_{\rm det,L}=\eta_L-\lambda_L \hat{O}_L i (c_{Q,L} c^\dagger_{P,L} - c_{P,L} c^\dagger_{Q,L})$, 
with
\begin{align}
\eta_L = &\left (|w_{1,Q}|^2 + |w_{3,Q}|^2 + |w_{3,P}|^2 + |w_{5,P}|^2 \right)/2 ,\\
\lambda_{L} =& [({\rm Re}(w_{1,Q}w_{3,P}^*))^{2} + ({\rm Re}(w_{3,Q}w_{5,P}^*))^{2} +({\rm Re}(w_{1,Q}w_{5,P}^*))^{2}]^{1/2} ,\\
\hat{O}_L= & -i \left[ ({\rm Re}(w_{1,Q}w_{3,P}^*)) \gamma_1 \gamma_3 + ({\rm Re}(w_{3,Q}w_{5,P}^*))\gamma_3 \gamma_5 +({\rm Re}(w_{1,Q}w_{5,P}^*))\gamma_5 \gamma_1 \right]/\lambda_L. \label{eq:operatore}
\end{align}
In fact $\hat{O}_{L}$ takes the form of Eq. (1) of the manuscript with $\cos\theta_{L}= {\rm Re}\left(w_{1,Q}w_{3,P}^*\right)/\lambda_L$,
$\sin\theta_{L}\cos\phi_{L}=-{\rm Re}\left(w_{3,Q}w_{5,P}^*\right) /\lambda_L$,
$\sin\theta_{L}\sin\phi_{L}={\rm Re}\left(w_{5,P}w_{1,Q}^*\right) /\lambda_L$.

The specific form of $H_{\rm det,L}^2$, which involves a single operator, allows to evaluate the series in Eq. \eqref{mela},  yielding
\begin{equation}
\vert \Psi_{L,(1,0)} \rangle = \left( c_{Q,L}^\dagger c_{Q,L} \right ) \left( \eta_{L}(\Delta t) -\lambda_L(\Delta t) \hat{O}_L i (c_{Q,L} c^\dagger_{P,L} - c_{P,L} c^\dagger_{Q,L})\right)(q_L c_{Q,L}^\dagger + p_L c_{P,L}^\dagger)  \vert 0\rangle_L 
\end{equation}
where the time functions $\eta_L$ and $\lambda_L$ are expressed by the series
\begin{align}
\eta_L(t) &= \sum_{p=0}^{\infty} \sum_{s=0}^p \left[ \frac{t^{4p}}{(4p)!} \frac{(2p)!}{(2s)! (2p-2s)!} \eta_L ^{2p-2s} \lambda_L^{2s} - \frac{t^{4p+2}}{(4p+2)!} \frac{(2p+1)!}{(2s)! (2p+1-2s)!} \eta_L ^{2p+1-2s} \lambda_L^{2s}\right], \\
\lambda_L(t) &= \sum_{p=0}^{\infty} \left[  -\frac{\lambda_L^{2p+1} t^{4p+2}}{(4p+2)!}+\sum_{s=0}^{p-1} \left[ \frac{t^{4p}}{(4p)!} \frac{(2p)! \eta_L ^{2p-2s-1} \lambda_L^{2s+1} }{(2s+1)! (2p-2s-1)!} - \frac{t^{4p+2}}{(4p+2)!} \frac{(2p+1)! \eta_L ^{2p-2s} \lambda_L^{2s+1}}{(2s+1)! (2p-2s)!} \right] \right].
\end{align}
This leads to the expression of the Kraus operator
\begin{gather}
M_{L,(1,0)} = q_L \eta_L(\Delta t) +i  p_L \lambda_L (\Delta t) \hat{O}_L. \label{eq:kraus}
\end{gather}
The result for $M_{R,(1,0)}$ is identical upon replacing $L \to R$. We can finally write the probabilities of interest as
\begin{align}
P^L_{(1,0)} &= |q_L|^2 \eta_L^2 +|p_L|^2 \lambda_L^2 +2{\rm Im} \left( q_Lp_L^*\right)\eta_L \lambda_L \langle \hat{O}_L\rangle, \label{eq:finalina1}\\
P^R_{(1,0)} &= |q_R|^2 \eta_R^2 +|p_R|^2 \lambda_R^2 +2{\rm Im} \left( q_Rp_R^*\right)\eta_R \lambda_R \langle \hat{O}_L\rangle, \label{eq:finalina2}\\
P_{(1,0,1,0)} &=  \left( \eta_L^2(\Delta t) |q_L|^2 + \lambda_L^2(\Delta t)  |p_L|^2 \right) \left( \eta_R^2(\Delta t) |q_R|^2 + \lambda_R^2(\Delta t) |p_R|^2 \right) \nonumber \\
& + 2 \langle \hat{O}_L \rangle  {\rm Im} (q_L p_L^*) \eta_L(\Delta t) \lambda_L (\Delta t) \left( \eta_R^2(\Delta t) |q_R|^2 + \lambda_R^2(\Delta t) |p_R|^2 \right) \nonumber \\
& + 2 \langle \hat{O}_R \rangle  {\rm Im} (q_R p_R^*) \eta_R(\Delta t) \lambda_R (\Delta t) \left( \eta_L^2(\Delta t) |q_L|^2 + \lambda_L^2(\Delta t) |p_L|^2 \right) \nonumber \\
& - 4 \langle \hat{O}_L \hat{O}_R \rangle \eta_L  (\Delta t) \eta_R (\Delta t) \lambda_L (\Delta t)\lambda_R (\Delta t) \, {\rm Im} (q_R p_R^*) {\rm Im} (q_L p_L^*), \label{eq:finalina}
\end{align}
where 
\begin{align}
\hat{O}_R= & -i \left[ ({\rm Re}(w_{2,Q}w_{4,P}^*)) \gamma_2 \gamma_4 + ({\rm Re}(w_{4,Q}w_{6,P}^*))\gamma_4 \gamma_6 +({\rm Re}(w_{2,Q}w_{6,P}^*))\gamma_6 \gamma_2 \right]/\lambda_R, \label{eq:operatore-d}\\
\eta_R = &\left (|w_{2,Q}|^2 + |w_{4,Q}|^2 + |w_{4,P}|^2 + |w_{6,P}|^2 \right)/2 ,\\
\lambda_{R} =& [({\rm Re}(w_{2,Q}w_{4,P}^*))^{2} + ({\rm Re}(w_{4,Q}w_{6,P}^*))^{2} +({\rm Re}(w_{2,Q}w_{6,P}^*))^{2}]^{1/2} 
\end{align}
The limit of small time $\Delta t$ or small couplings, $\lambda_L$, $\lambda_R$, $\eta_L$, $\eta_R$, which correspond to the weak measurement regime, is the one discussed in the manuscript (with the further simplifications due to of $w_{\alpha,j} \in \mathbb{R}$ and $p_L=-iq_L$). 

Based on the equations above, the entanglement detection protocol consists in a series of measurements of charge transition rates among the dots. It is assumed that all the measurement calibration parameters ($\lambda_L$,$\lambda_R$, $\eta_L$, $\eta_R$, $p_L$, $q_L$, $p_R$, $q_R$) are known. From Eqs. \eqref{eq:finalina1} and \eqref{eq:finalina2}, measurements of $P^L_{(1,0)}$ and $P^R_{(1,0)}$ can be used to determine $\langle \hat{O}_L\rangle$, $\langle \hat{O}_R\rangle$, so that the only unknown element of \eqref{eq:finalina} is $\langle \hat{O}_L \hat{O}_R \rangle$, and it can be therefore determined by a measurement of $P_{(1,0,1,0)}$. Having established a procedure to measure correlations between left and right operators, one can detect the violation of CHSH inequalities due to entangled states.
In the example considered in the manuscript, in which the state of the Majorana system is given by $A=D=0$, one has a maximally entangled state in the $135|246$ partition. 
In order to obtain a corresponding maximal violation of the CHSH inequalities, $\mathcal{C}_{135|246}=2\sqrt{2}$, the required operators are  $\hat{O}_L=\hat{Z}_L$, $\hat{O}_L'=\hat{X}_L$, $\hat{O}_R=(\hat{Z}_R+\hat{X}_R)/\sqrt{2}$, $\hat{O}_R'=(\hat{Z}_R-\hat{X}_R)/\sqrt{2}$. They are obtained by properly tuning the parameters in Eq.\eqref{eq:operatore}  and Eq.\eqref{eq:operatore-d}. Specifically, by tuning $w_{5,P}=0$, one obtains, from Eq. \eqref{eq:operatore}, $\hat{O}_L=-i \gamma_1 \gamma_3=\hat{Z}_L$; similarly  a measurement of $\hat{O}'_L=\hat{X}_L$ is realized by tuning $w_{1,Q}=0$. In the right site, one sets $w_{2,Q}=-w_{6,P}\ll w_{4,Q}=w_{4,P}$ to obtain $\hat{O}_R=\frac{1}{\sqrt{2}}[(1+\mathcal{O}(\epsilon))\hat{Z}_R+(1+\mathcal{O}(\epsilon))\hat{X}_R+\mathcal{O}(\epsilon)\hat{Y}_R]$ and $w_{2,Q}=w_{6,P}\ll w_{4,Q}=w_{4,P}$ to obtain $\hat{O}'_R=\frac{1}{\sqrt{2}}[(1+\mathcal{O}(\epsilon))\hat{Z}_R-(1+\mathcal{O}(\epsilon))\hat{X}_R+\mathcal{O}(\epsilon)\hat{Y}_R]$, where $\epsilon=w_{2,Q}/w_{4,Q} \ll 1$ can be independently tuned to be arbitrarily small. These are the operators required to have a maximal violation of the CHSH inequalities, as reported in the manuscript.

It is interesting to note that for long time and/or strong coupling, the measurement does not approach a strong projective measurement. This can be seen from Eq. \eqref{eq:kraus}, which generically is not a projector (e.g. $M_{L,(1,0)}^2 \neq M_{L,(1,0)}$). This feature is due to the finite dimension of the HIlbert space of the quantum dots used as an ancilla in the detection scheme. In this case the measurement time and coupling constants enter the Kraus operator through the superposition of a finite number of of periodic functions. This fact can be seen explicitly from Eq. \eqref{eq:grande}, where $\Delta t$ appears as a factor the exponential of a finite-dimensional matrix.

Importantly, Eq. (8) shows that a measurement realized  with the proposed scheme leads generically to a probability of the form Eq. (8) in the manuscript regardless of the duration and coupling strength. The proposed scheme can be therefore used as a test for CHSH inequalities even beyond the weak coupling limit. However, for intermediate measurement strength, the detector calibration requires a fine-tuning of the coupling time. 
This has to be compared with the weak measurement regime where a power of $\Delta t$ appears as  a common prefactor of the detector signal. Therefore, although the proposed measurement scheme works for arbitrarily strength of the measurements and coupling time, the weak measurement regime offers a more general form of the detector's signal, which, in turns, reduces the need for fine tuning in a given experimental setup.

\section{Calculation of the minimal violation of the CHSH inequality}

In the manuscript we have introduced the minimal violation of CHSH inequalities,  $\mathcal{C}  =\min_{\vert\psi\rangle}\left\{ \mathcal{C}_{0}(\vert\psi\rangle)\right\} \approx 2.078$, where 
$\mathcal{C}_{0}(\vert\psi\rangle)  \equiv \max\left\{ \mathcal{C}_{135|246}(\vert\psi\rangle),\,\mathcal{C}_{564|132}(\vert\psi\rangle),\,\mathcal{C}_{421|563}(\vert\psi\rangle),\,\mathcal{C}_{641|352}(\vert\psi\rangle)\right\} $ . We report here the details of the calculation of $\mathcal{C}$. 

Let us first introduce, for notational convenience, 
\begin{align*}
C_{1}  \equiv4\vert AD-BC\vert^{2}, \, C_{5}  \equiv4\vert AC-BD\vert^{2}, \, 
C_{4}  \equiv4\vert AB-CD\vert^{2}, \, C_{6}  \equiv\vert A^{2}-B^{2}+C^{2}+D^{2}\vert^{2}.
\end{align*}
We are looking for a state, $\vert \psi_0 \rangle$, that will minimize $\mathcal{C}_0$, with the goal of showing that even the minimal value of $\mathcal{C}_0$ exceeds the classical bound of CHSH inequalities. Since $\mathcal{C}_{135|246}=2\sqrt{1+C_{1}}$, $\mathcal{C}_{564|132}=2\sqrt{1+C_{5}}$,
$\mathcal{C}_{421|563}=2\sqrt{1+C_{4}}$, $\mathcal{C}_{641|352}=2\sqrt{1+C_{6}}$,
depend monotonically on the newly defined quantities,  $\vert\psi_{0}\rangle$
is obtained by minimizing
\[
\mathcal{K}=\min_{\vert\psi\rangle}\left\{ \max\left\{ C_{1}(\vert\psi\rangle),\,C_{5}(\vert\psi\rangle),\,C_{4}(\vert\psi\rangle),\,C_{6}(\vert\psi\rangle)\right\} \right\} .
\]
We introduce also the convenient parametrization
\begin{equation}
A=a\,e^{i\alpha},\:B=b\,e^{i\beta},\:C=c\,e^{i\gamma},\;D=d\,e^{i\delta},\label{eq:parametri}
\end{equation}
where $a,b,c,d\in\mathbb{R}^{+}$, $\alpha,\beta,\gamma,\delta\in[0,2\pi)$,
and 
\begin{equation}
a^{2}+b^{2}+c^{2}+d^{2}=1.\label{eq:norma}
\end{equation}
A direct calculation shows that  the stationary points of $C_{1}$ with the constraint (\ref{eq:norma}) correspond to the global maximum, $C_{1}=1$, or the global minimum, $C_1=0$. The minimum $C_1=0$ is obtained for the set of parameters where $ad=bc$, and $\alpha+\delta-\gamma-\beta=2n\pi$. Repeating the search for extremal points of $C_{5}$, $C_{4}$, $C_{6}$ yields similar sets since the latter functions are all obtained from $C_1$ by unitary transformations of the vector $(A,B,C,D)^T$. Since the intersection of all these sets of points is empty (i.e. $C_1$, $C_5$, $C_4$, $C_6$ cannot be all simultaneously vanishing as stated in the manuscript), the state, $\vert\psi_{0}\rangle$, which minimizes $\mathcal{K}$ must fulfil the condition $C_i(\vert\psi_{0}\rangle)=C_j(\vert\psi_{0}\rangle)$, $i \neq j$.

A direct calculation shows that the stationary points of $C_1$ with the conditions \eqref{eq:norma} and 
\begin{equation}
C_{1}=C_{5} \label{eq:uguaglianza}
\end{equation}
are the same global minima and maxima obtained without the constraint\eqref{eq:uguaglianza}. Iterating the argument for $C_{1}=C_{5}=C_{4}$ one concludes that $\vert\psi_{0}\rangle$ is constrained to the surface defined by
\begin{equation}
C_{1}=C_{5}=C_{4}=C_{6}.\label{eq:condizione}
\end{equation}
The minimum of $C_{1}$ under the condition Eq. (\ref{eq:condizione})
is not a global minimum. In fact, at most three of these functions
can be simultaneously vanishing. Specifically, the set of points where $C_{1}=C_{5}=C_{4}=0$ is 
\begin{equation*}
\{ A=B=C=D \} \cup \{A=-C=B=-D\} \cup \{A=-B=C=-D\} \cup \{A=-B=D=-C \}
\end{equation*}
For all these points $C_{6}\neq0$. 

In the following, to determine the minimum of $C_1$ under the condition in (\ref{eq:condizione}) and \eqref{eq:norma}, we first reparametrize the states constrained to $C_{1}=C_{5}=C_{4}$, and then impose the further condition $C_{1}=C_{6}$.
As a preliminary, starting from the parameters in Eq. (\ref{eq:parametri}),
we assume, without loss of generality,  $\delta=0$, since the latter can be
reabsorbed in the overall (non-physical) phase of the state $\vert\psi\rangle$.
We then have
\begin{gather}
C_{1}=C_{5}\implies(a^{2}-b^{2})(d^{2}-c^{2})=4\,abcd\sin(\alpha-\beta)\sin(\gamma), \\
C_{1}=C_{4}\implies(a^{2}-c^{2})(d^{2}-b^{2})=4\,abcd\sin(\alpha-\gamma)\sin(\beta), \\
C_{5}=C_{4}\implies(a^{2}-d^{2})(c^{2}-b^{2})=4\,abcd\sin(\alpha)\sin(\beta-\gamma),
\end{gather}
where the last equation is a redundant statement of the first two.
To analyze the above constraints let us, as a first attempt, assume $\sin(\alpha-\beta)\sin(\gamma)\neq0$, $\sin(\alpha-\gamma)\sin(\beta)\neq0$,
$\sin(\alpha)\sin(\beta-\gamma)\neq0$; the only solutions of the
above equations are
\begin{align}
a & =b=c\neq d\label{eq:sola}\\
a & =b=d\neq c\label{eq:solb}\\
b & =c=d\neq a\label{eq:solc}\\
a & =c=d\neq b\label{eq:sold}
\end{align}
Each of these equations, together with the normalization in Eq. \eqref{eq:norma} allows us to eliminate three parameters.
Each of these solutions (Eqs. (\ref{eq:sola},\ref{eq:solb},\ref{eq:solc},\ref{eq:sold})) implies 
\begin{equation}
abcd\sin(\alpha-\beta)\sin(\gamma)=abcd\sin(\alpha-\gamma)\sin(\beta)=abcd\sin(\alpha)\sin(\beta-\gamma)=0\label{eq:condizione-intermedia}
\end{equation}
so either $a=0$, or $b=0$, or $c=0$, or $d=0,$ or, contrary to our assumption,
\begin{equation}
\sin(\alpha-\beta)\sin(\gamma)=\sin(\alpha-\gamma)\sin(\beta)=\sin(\alpha)\sin(\beta-\gamma)=0.\label{eq:seni}
\end{equation}
%The latter, together with Eqs. (\ref{eq:sola},\ref{eq:solb},\ref{eq:solc},\ref{eq:sold})
%fulfills $C_{1}=C_{5}=C_{4}$. 
The condition $a=0$, together with
Eqs.(\ref{eq:condizione},\ref{eq:sola},\ref{eq:solb},\ref{eq:solc},\ref{eq:sold})
necessarily leads to $C_{1}=C_{5}=C_{4}\neq C_{6}$, and is therefore
excluded. The same argument excludes $b=0,\,c=0,\,d=0$. We thus need to abandon our earlier assumption. 
Eq. (\ref{eq:seni}), instead, has the different possible solutions:
\begin{align}
&  \alpha=n\pi,\,\beta=m\pi,\,\gamma\in [0,2\pi), \label{eq:sol11} \\
&  \alpha=n\pi,\,\gamma=m\pi,\,\beta\in[0,2\pi), \label{eq:sol12} \\
&  \beta=n\pi,\,\gamma=m\pi,\,\alpha\in[0,2\pi), \vee \, (\beta-\alpha=n\pi,\,\gamma-\alpha=m\pi,\,\alpha\in[0,2\pi)),\label{eq:sol13}
\end{align}
where $n,m\in \mathbb{Z}$. As a result he surface defined by $C_1=C_5=C_4$ consists of the combinations of each of Eqs. (\ref{eq:sola},\ref{eq:solb},\ref{eq:solc},\ref{eq:sold}) wth each of the Eqs. (\ref{eq:sol11},\ref{eq:sol12},\ref{eq:sol13}).

Let us consider all these different cases separately. Using Eq. (\ref{eq:sola}) and Eq. \eqref{eq:sol13} in the definitions of $C_1$ and $C_6$ yields 
\begin{align}
C_{1,\pm}  =4a^{2}\left(1-2a^{2}\pm2a\sqrt{1-3a^{2}}\cos(\alpha)\right),\, C_{6}  =\vert(1-3a^{2})+a^{2}e^{2i\alpha}\vert^{2}, \label{min1}
\end{align}
where $C_1$ has two different expressions $C_{1,\pm}$ corresponding to even and odd parities of $m \cdot n$ respectively. Combining Eq. \eqref{eq:sola} with Eq. \eqref{eq:sol11} yields 
\begin{align}
C_{1,\pm}  =4a^{2}\left(1-2a^{2}\pm2a\sqrt{1-3a^{2}}\cos(\gamma)\right),\, C_{6}  =\vert(1-3a^{2})+a^{2}e^{2i\gamma}\vert^{2}, \label{min2}
\end{align}
and using Eq. \eqref{eq:sola} and \eqref{eq:sol12} for $C_1$ and $C_6$ results in
\begin{align}
C_{1,\pm}  =4a^{2}\left(1-2a^{2}\pm2a\sqrt{1-3a^{2}}\cos(\beta)\right),\, C_{6}  =\vert(1-a^{2})-a^{2}e^{2i\beta}\vert^{2}. \label{min3}
\end{align}
Combing Eq. (\ref{eq:solb}) and Eq. (\ref{eq:solc}) with each of the Eqs. (\ref{eq:sol11},\ref{eq:sol12},\ref{eq:sol13}) leads to equations identical to (\ref{min1},\ref{min2},\ref{min3}).
Repeating the procedure for the case of Eq. (\ref{eq:sold}) results instead in different expressions for $C_6$. Specifically, combining Eq. (\ref{eq:sold})  with each of the Eqs. (\ref{eq:sol11},\ref{eq:sol12},\ref{eq:sol13}) in the definition of $C_1$ and $C_6$ gives
\begin{align}
C_{1,\pm}  =4a^{2}\left(1-2a^{2}\pm2a\sqrt{1-3a^{2}}\cos(\alpha)\right),\, C_{6}  =\vert(5a^{2}-1)+a^{2}e^{2i\alpha}\vert^{2}, \label{min4}
\end{align}
or
\begin{align}
C_{1,\pm}  =4a^{2}\left(1-2a^{2}\pm2a\sqrt{1-3a^{2}}\cos(\gamma)\right),\, C_{6} =\vert(5a^{2}-1)+a^{2}e^{2i\gamma}\vert^{2}, \label{min5}
\end{align}
or
\begin{align}
C_{1,\pm}  =4a^{2}\left(1-2a^{2}\pm2a\sqrt{1-3a^{2}}\cos(\beta)\right),\, C_{6}  =\vert(3a^{2}-1)+3a^{2}e^{2i\beta}\vert^{2}. \label{min6}
\end{align}

For each of the cases in Eqs. (\ref{min1},\ref{min2},\ref{min3},\ref{min4},\ref{min5},\ref{min6}), we need to minimize $C_1$ with $C_1=C_6$ and the further constraints $0\leqslant a \leqslant 1/\sqrt{3}$, $\alpha, \beta, \gamma \in [0,2\pi)$. We are interested in the global minimum of $C_1$ over all these different cases.
Note that Eq.(\ref{min2}) and Eq.(\ref{min1}) are identical upon reparametrizing $\gamma \to \alpha$. The same holds for Eq.(\ref{min4}) and Eq.(\ref{min5}). Therefore we are left with the following eight minimization problems for $C_{1\pm}$ over the parameters $0\leqslant a \leqslant 1/\sqrt{3}$ and $\alpha, \beta, \gamma \in [0,2\pi)$ with the
condition $C_{1}=C_{6}$:
\begin{align}
C_{1,\pm}  =4a^{2}\left(1-2a^{2}\pm2a\sqrt{1-3a^{2}}\cos(\alpha)\right),\label{eq:f1}\, C_{6}  =\vert(1-3a^{2})+a^{2}e^{2i\alpha}\vert^{2},
\end{align}
or
\begin{align}
C_{1,\pm}  =4a^{2}\left(1-2a^{2}\pm2a\sqrt{1-3a^{2}}\cos(\beta)\right),\label{eq:f2}\, C_{6}  =\vert(1-a^{2})-a^{2}e^{2i\beta}\vert^{2},
\end{align}
or
\begin{align}
C_{1,\pm}  =4a^{2}\left(1-2a^{2}\pm2a\sqrt{1-3a^{2}}\cos(\alpha)\right),\label{eq:f3}\,
C_{6}  =\vert(5a^{2}-1)+a^{2}e^{2i\alpha}\vert^{2},
\end{align}
or
\begin{align}
C_{1,\pm} =4a^{2}\left(1-2a^{2}\pm2a\sqrt{1-3a^{2}}\cos(\beta)\right),\label{eq:f4}\,
C_{6}  =\vert(3a^{2}-1)+3a^{2}e^{2i\beta}\vert^{2}.
\end{align}
From numerical calculations, we obtain for the lowest minimum in the above cases, $\min_{a,\alpha} C_1 \approx 0.031$, hence $\mathcal{C} \approx 2.031$.
%For the Eqs. (\ref{eq:f1}) The condition $C_{1}=C_{4}$ allows us
%to eliminate $\cos(\alpha)$ form the equations, yielding 
%\[
%\cos(\alpha)=\frac{a\sqrt{1-3a^{2}}\left(4a^{2}\pm\sqrt{2(-16a^{4}+12a^{2}-1)}\right)}{2a^{2}(1-3a^{2})},
%\]
%and 
%\[
%C_{1\pm,\pm}(a)=4a^{2}\left(1-2a^{2}\pm4a^{2}\pm\sqrt{-2((-16a^{4}+12a^{2}-1))}\right).
%\]
% The minimum of the four different $C_{1\pm,\pm}(a)$ in the interval
%defined by $-1<\cos(\alpha)<1$, yields the numerical value $\min\{C_{1}=C_{5}=C_{4}=C_{6}\}=0.080$.
%Repeating the same procedure for the problem defined by Eqs. (\ref{eq:f1})
%%, yields $\min\{C_{1}\}=0.163$

\end{document}